# Review for future research in digital leadership

Raluca Alexandra Stana[1], Louise Harder Fischer[2], and Hanne Westh Nicolajsen[3]

[1,3] IT University of Copenhagen, Copenhagen, Denmark
[2] Copenhagen Business School, Copenhagen, Denmark

**Abstract**. Information Technology (IT) enables challenges and opportunities for how enterprises organize themselves and how work unfolds in digital settings. The changes in technology, work, organizations, and humans' mindset, call for new ways to discuss leadership. In these new digital settings, traditional leadership no longer holds. Instead, new forms of digital leadership are needed. In this paper we review the IS research perspective on digital leadership. We conduct a systematic literature review combined with a hermeneutical approach. We compare the findings from our literature review with a theoretical lens inspired by e-leadership. We find that IS research has an exclusive scope on strategic implications, leaving out topics such as followership and emotions. On the other hand, IS research contributes with exactly these issues of strategic leadership and business transformation to the digital leadership discussion. We contribute to IS research by defining digital leadership, proposing a theoretical lens of three levels of analysis of digital leadership, as well as paths for future research.

## 1      Introduction

Scholars argue that leadership plays a critical role in the introduction and adoption of digital change initiatives in organizations [1], [2]. At the same time, leadership is also impacted by the introduction and usage of technology [3], as technology is changing both the context for leadership [4], as well as how leadership can be enacted in the new context [5], [6].

With the increase focus on business transformation, IT requires commitment starting from the CEO and top management team [7]. As Peppard and Ward discuss, we are moving from an era of building Information System (IS) strategy, to an era where IS capability is needed throughout the entire organization, including top management [8].

Technology institutionalizes workflows, and the employees who used to perform routine tasks are increasingly freed by communication and collaborative technology to focus on more creative and complex tasks [9], [10], [11]. Due to the abundance of information and technology, formerly constrained and restricted in the format of the hierarchy, non-routine discretionary work is increasingly self-organized [11] and work takes place around smaller cross-hierarchical networks [1], [12]. The traditional role of the manager, as direct supervisor, controller and coordinator is changing. For example, enterprise social networks (ESN) enable employees to self-organize around smaller digital networks of productive and competent peers and withdraw from the larger socialization context of the organization [13]. These networks can disrupt the leadership platform, allowing for leadership to emerge from followers, dyads, context, or the collective [5], [6].

However, hierarchy disruption is only one out of the many disruptions of the leadership platform. While work is becoming more complex [11], it has also changed and evolved from the temporal, spatial, and task defined job [14], to employees being able to perform their work anytime or anywhere [15] [16]. Work is also more autonomous and innovative [17] and a greater part of the workforce is contemporary employed (e.g. freelancers). In the post bureaucratic and digital organization, productivity, formerly a responsibility of the manager, is now in the hands of the autonomous knowledge professional [17].

Digital communication also poses challenges on leadership. Avolio et al suggest that digital communication affects how one's emotions are perceived [18]. For example, in digital communication receivers may perceive a message as less positive than the leader intended it to be [19]. As traditional leadership theories are built on the assumption of face-to-face interactions [4], and face-to-face interactions cannot be directly translated to digital interactions [19] [20] [21] [22], there is a need to reconsider the applicability of traditional leadership theories in the digital organization.

Besides the changing fabric of the organization, the changing mindset of employees also has an impact on leadership. Bass argue that the confirming worker of the 1950s, has been replaced with the sceptical worker of the 1990s [23]. Moreover, Avolio et al highlight that the "Millennial" generation have believe that leaders should serve rather than direct [18].

In sum, the changes in technology, work, organizations, and humans' mindset, call for new ways to discuss leadership. Considering the critical role that leadership plays in tackling organizational and technological changes, we ask: What is the current understanding of digital leadership in IS research and how can it inform future IS research?

We acknowledge our pre-supposition before entering the literature review: leadership in the domain of IS has so far been narrowly addressed in terms of IT-leadership and virtual leadership, in which IT leadership often is limited to either an IT department, the IT strategy, or how to manage larger ERP-implementations [24], while virtual leadership is mostly focused on the management of virtual teams [25] [26].

The trajectory of e-leadership from the domain of leadership and psychology [4] [18] [27] exhibits a broader focus of opportunities and challenges that leaders are confronted with in the digital era of work. Thus, we compare the concepts emerged from our systematic and hermeneutical literature review on digital leadership in IS, with a theoretical lens built around central constructs and levels inspired by e-leadership [18]. We contribute to IS research with a definition of digital leadership, a theoretical lens, and by defining future paths of research.

The paper is structured as follows: in section two we describe digital leadership and the theoretical lens, section three describes the methodology, in section four we analyse, in section five we discuss, we conclude in section six, while section seven presents our limitations.

## 2      Digital leadership and theoretical lens

Before we enter the body of knowledge from the literature search, we build an understanding of digital leadership, and use it as an unlocking device [28] to start the literature review. We build our definition of digital leadership departing from the definition of e-leadership and enhance it with Sørensen's interpretation of technology within IS [24].

### 2.1      E-leadership

Some of the most influential and recent work on e-leadership is found within the field of leadership and psychology. In building the theoretical lens that we later use to compare our literature review on digital leadership in IS, we lean on these contributions.

The first paper on e-leadership, defines it as a process of social influence mediated by Advanced Information Technology (AIT) which can occur at any level in an organization, and it can be one-to-one or one-to-many [4] . This implies a multilevel approach and a three-directional influential interaction between leaders, followers, and AIT within organizations. This definition was later expanded by Avolio et al to consider how e-leadership is influencing the appropriation of AIT, as well as how AIT impacts e-leadership [18].

Advanced Information Technology (AIT) is defined as tools and techniques designed for the participation of employees in collection, processing, management, retrieval, transmission, and display of data and knowledge [4].

### 2.2      Digital Leadership

We define digital leadership as a process of social influence mediated by technology to produce a change in attitudes, feelings, thinking, behaviour and/or performance with individuals, groups, and/or organizations, which can occur at any hierarchical level in an organization and can involve one-to-one and/or one-to-many interactions

We have decided to use the term digital leadership for three reasons:

Firstly, the definition of e-leadership is based on AIT, whereas we would like to define digital leadership by drawing on Sørensen's understanding of technology within IS. Sørensen is encouraging IS scholars to move their focus beyond the analysis of one organization and one major technology embedded within (e.g. Enterprise Resource Planning (ERP) systems), towards understanding the nascent socio-technical configurations emerging from innovation and human actions [24]. More precisely, Sørensen is suggesting a few trends to be more closely considered: digitalization, distribution, and scale, and their implications in organizations, humans, and work. Avolio's et al definition of AIT [4], combined with Sørensen's description of the IS perspective on technology [24], provide a more complete picture of the technological challenges in the post-bureaucratic and modern organizations, which are disrupting the leadership system. This is the view on technology that we employ when discussing digital leadership.

Secondly, by using the term digital leadership we wish to differentiate from leadership and leadership in the digital age. Leadership as a field of research draws on a tradition of face to face interactions and builds on a body of knowledge and traditions older than e-leadership.

Thirdly, the usage of the term "digital leadership" bring us closer to practitioners' concerns [29]. Our initial keyword search (January, 2018) on Google Scholar on "digital leadership" uncovered a number of 1360 titles, out of which most of them were materials for practitioners that contained "digital leadership" in their title.

We would also like to note that during our literature review we did not find a satisfactory definition of digital leadership within IS, suggesting that this field is emerging. The nascent aspect of digital leadership within IS motivates us to turn to a theoretical lens built from e-leadership.

### 2.3 Theoretical Lens

We build our theoretical lens on digital leadership on the work on e-leadership by Avolio et al, who have identified three levels of e-leadership: macro, micro, and meso in their literature review [18]. In this chapter we describe the three levels of digital leadership. Later, these levels will be used to analyse the findings from our literature review on digital leadership within IS.

**Macro level.** The macro level addresses the strategic implications of digital leadership in organizational change and transformation, and how it influences the implementation, adoption, or adaptation of technology at a strategic level.

**Micro level**. Digital leadership can emerge from multiple levels, such as individuals assuming the role of the leader, from the leader-follower dyads, from followers, from the context, or from the collective. Digital leadership can be transmitted in four ways: through one's traits (how one is), behaviours (what one does), cognitions (what and how one thinks), or emotions (what one feels).

**Meso level.** Digital leadership at a meso level is concerned with changes within the work context, such as use of information technology and social networks, and how they influence leadership. For example, how increased transparency and openness influences hierarchies, and its impact on leadership.

## 3 Methodology

In this section we explain the search strategy that we used to arrive at the 10 relevant IS articles. We apply the Webster and Watson's literature review techniques [18] while also drawing upon the hermeneutic approach as represented by Boel and Cesec-Komanoviz [30]. This implies a methodology that involves circular interpretation, and a process that aims for saturation in understanding how digital leadership is viewed in IS.

### 3.1 Search strategy

During January and February 2018, we performed multiple searches in the senior basket of IS journals, SpringerLink, Google Scholar, AIS, and ABI Inform for the terms "digital leadership", "e-leadership", "virtual leadership", and "IT leadership". The criteria to deem articles relevant for our literature review was the content of the title and the outlet. As our research question relates to IS research, we only considered IS journals and conferences as relevant. We have included titles that contained the aforementioned keywords, as well as titles pertaining to leadership and technology. We ended up with a number of 17 articles.

The first analytical layer was weaving out articles that were not relevant, from the ones found after the procedure described above. From the preliminary number of 17 articles found during our search, we excluded articles that belonged to a specific domain (e.g. healthcare or education), and articles that focused on digital leadership as a competitive advantage. The final number of articles after these procedures was six.

Next, we found four more relevant articles that were cited extensively in the first six articles. We started reading the first six articles and then the four following articles and mapped the emerging concepts. We see these first ten articles as a pilot for a more extended review.

## 4. Analysis

In this chapter we analyse the articles discovered through our literature review. We start with a few general comments, then we analyse the contextual data of the articles, and after, we categorize and discuss the papers based on the theoretical lens described in chapter 2.3.

**4.1. Focus and theoretical background**

We observe a focus on the changing role of the CIO (Chief Information Officer) as a result of the increased focus on information technology in organizations [31] [32] [33] [34] [35]. This changing role is explored from several perspectives: the means of the CIO to influence employees on the business side in terms of IT capabilities [32]; the individual and organizational factors characterizing the CIO position and their influence on strategic IT alignment [31]; the supply-side and the demand-side leadership of the CIO and the outcomes of such leadership [33]; the changing role of the CIO in organizations [34]; or the changing responsibility of the CIO who now needs to communicate technological opportunities to the board of directors [35].

The analysis of the changing role of the CIO is rooted in Mintzberg's managerial roles [32] [34], Strategic Alignment Theory [31], or exploration/exploitation theory [33].

The board of directors' implications in IT is also a focus, either in terms of the involvement needed in decision-making and planning [35], or in terms of the capabilities and competencies needed to lead Enterprise Business Technology Governance (EGBT) [29].

Top management support (TMS) is deemed to be an important factor when it comes to major technological transformations in organizations from several perspectives: Enterprise Systems (ES) life cycle and the type of leadership needed in each of the ES phases [36]; the competencies required for the management function in order to reach value creation from IT investments, the type of leadership required for IT-enabled transformations, and the role of the management to bring about the necessary changes [37]; and the importance of TMS in resolving conflicts, reinforce norms, or unfreezing institutional routines [36].

Alignment between business strategy and digital technology is also a focus, in particular how to derive alignment between business needs and technology innovation [38], or how the CIO can influence this alignment [31]. In both instances, Strategic Alignment Theory is being employed as a means to explore the optimal connection between business and IT.

**4.2. Data and demographics**

We notice a focus on executives or organizations based in the US [31] [33] [34] [35]; an article focused on Europe [38]; an article focused on China [36]; while other papers do not specify the location [39] [29] [37] [32].

The method for data collection varies from interviews [36] [37] [38], to surveys [29] [31] [32], or case studies [36].

We observed that in all the papers where the gender was specified, a high percentage of respondents were male: 78% [29], 70% [31], and 61% [32]. At the same time, the age varies from between 46 and 65 amongst 73% of the respondents [29]; an average of 46.2 years of age [31]; between 36 and 55 years [32], or 49 years [34]. The respondents were in all cases executives, or decision-makers, except for one instance where employees were also interviewed [36]. However, the focus of the findings was on the executives, top management, or decision makers.

We also note that culture, country, gender, or age are not discussed in relation to the findings of any of the papers where these aspects are specified, while in the rest of the cases these are neither specified nor discussed.

### 4.3 Literature review through the theoretical lens

We have chosen to apply the theoretical lens of digital leadership described in chapter 2.3 to analyse the articles discovered during our literature search. In this section, the theory of Avolio et al is further explicated, as we use their concepts and sub-categories. This provides a structured approach to our analysis, summarized in Table 1.

Table 1. Theoretical lens applied on literature review analysis

| Macro | | Wunderlich and Beck [31]; Wunderlich and Beck [32]; Chen et al [33]; Shao et al [36]; Agarwal et al [37]; Li et al [38]; Peppard et al [39]. | | | |
|---|---|---|---|---|---|
| Micro | | Traits | Behavior | Cognition | Emotions |
| | Leader | Shao et al [36]; Agarwal et al [37]; | Wunderlich and Beck [32]; Grover et al [34]; Andriole [35]; Shao et al [36]; Agarwal et al [37]; Li et al [38]; | Valentine and Stewart [29]; Grover et al [34]; Andriole [35]; Shao et al [36]; Agarwal et al [37]; Li et al [38]; | |
| | Followers | | | | Chen et al [33] |
| | Dyad | | | | |
| | Collective | | | | |

| | Context | | | | |
|---|---|---|---|---|---|
| Meso | | Valentine and Stewart [29]; Wunderlich and Beck [32]; Chen at al [33]; Grover et al [34]; Andriole [35]; Agarwal et al [37]; Li et al [38]; Peppard et al [39]. | | | |

**Macro**

We notice a high focus on transformational and transactional leadership styles, where transformational leadership is used as a device to advocate for a specific leadership style in times of technological changes within the organization [31] [33] [36] [37] [38].

Building on transformational and transactional leadership, the concept of leadership contingency is discussed in relation to the ES phase: transformational leadership is seen as more effective in the ES adoption phase, whereas transactional leadership is more effective in the ES implementation phase, and a combination of transformational and transactional is more effective in the ES assimilation and extension phase [36].

Supply-side and demand-side CIO leadership as concepts are based on the exploitation and exploration theories and are described as the CIO's capability to exploit the existing IT resources and competencies or to explore new IT-driven business opportunities that will lead to organizational innovations and business growth [33].

Another concept discovered at the macro level is strategic alignment between business and IT [31] [38]. Furthermore, Peppard et al discuss the strategic competencies needed at a macro level in order to fully leverage value from the IT investments of the organizations [39], while Chen et al point towards exploitation and exploration strategies in order to both leverage the present IT capabilities and seek new opportunities [33].

**Micro**

The articles in the literature review discuss the micro level exclusively from the perspective of the leader in an official leadership role and leader's behaviors and cognition are the most discussed dimmensions (see Table 1.). Traits are discussed less (see Table 1.), while emotions are only mentioned once and between the CIO and the CEO [33]. Dyads, Context, or Collective are not discussed at all, while Chen et al briefly mentions followers [33].

*Leader*
The leader in the position of the director, executive, top management, or CIO represents the focus of IS research at a micro level [31] [34] [35] [37] [38].

*Leader - Traits*
Charismatic traits of a leader are exemplified through the actions of the leader in relation to a case study [37].

*Leader - Cognition*
Leader cognition is touched upon mainly as a consequence of discussing IT leadership or the CIO's roles or style [31] [32] [38], but some papers [37] discuss it additionally to roles, in terms of skills and attitudes of the CEO. Valentine and Stewart argue about the competencies and capabilities of the board of directors in terms of their knowledge and ability to understand aspects of IT [29].

*Leader - Behaviour*
Behaviours are discussed at length, predominantly in relation to Mintzberg's managerial roles [31] [32] [34]. Valentine and Stewart talk about the ability of the leader in the board of directors to oversee IT governance or to evaluate risks, which is not an observed behaviour per se, but rather a recommendation of future action [29].

*Dyad*
Chen et al suggest that followers are more comfortable to challenge their leaders in small power distance cultures and that empowering followers with more structural power will increase the collaboration. They address this issue in the context of arguing for a smaller power distance between the CIO and the CEO, so although the dyad is addressed, the conversation takes place at the C-suit level [33].

*Context*
Chen et al discuss that the organizational context as well as the organizational culture have an influence on the CIO, as there are underlying power relationships and influence processes to be found in all organizations, that have a direct impact on CIO (e.g. smaller power distance reduces the emotional distance between leader and follower, making the follower feel comfortable to challenge the leader) [33].

**Meso**
The papers categorized in the meso level, mostly discuss leadership skills, roles, and capabilities that need to change or have changed due to IT.

Structural power is brought into conversation as the ability of the CIO to act as a leader, which is influenced by his or her individual capability, but also by organizational factors that can either facilitate or hinder his or her level of leadership [33]. The higher the structural power of the CIO, the higher the perceived decision-making authority of the IT leader in an organization, which influences the CIO's ability to transfer value-creating information and relevant insights towards top management [31]. The hierarchical level of management is a dimension that influences the roles of a leader, as at higher-level managerial roles the focus is on more external roles to bridge the organization with the outside environment, whereas at lower hierarchical levels, the focus is on internal responsibilities [34].

While some papers discuss hierarchies and structural power [31] [34], which is also a sub-dimension of the meso level, most papers classified as addressing the meso level are discussing the changing nature of leadership as a result of digitalization, business transformation, or the need for value creation with IT [29] [31] [34] [35] [37]. In one instance, the organization as a whole is subject to a competencies shift in order to support the digitalization of the business processes [39].

The relation between Leadership and Technology is mentioned in several papers: to point out the importance of having EBTG competencies among the board of directors [29]; to point out the role of the leadership in technology adoption [36], or innovation [37]; or to suggest that leadership and technology influence each other [38].

## 5. Discussion and Future paths of research

We find that the theoretical lens on e-leadership to be used to understand digital leadership in IS has strengths and weaknesses. The macro level could be further expanded to include strategic leadership (transactional and transformational leadership), strategic alignment, strategic competencies, leadership contingency, and exploration and exploitation strategies. In addition, IS research is concerned with discussing leadership roles, skills, responsibilities, competencies and styles, dimensions which could also be added to the micro level of the theoretical lens. The meso level could be enhanced with the concepts of digitalization, business transformation, and the need to create value with IT.

If this is the main body of knowledge on digital leadership within IS research, then we suggest the following paths of future research:

### 5.1     Emotions and Digital Leadership

The field of emotions hasn't been connected with digital leadership in IS so far, which can also be seen in Table 1. With the advances in understanding emotions, a new research path could be focused on how emotions influences leadership mediated by technology, and how emotions are transmitted, perceived or modified in digital communication and leadership. The field of Sociology of Emotions could bring the methodological and theoretical lenses needed for the field of digital leadership in IS to analyse emotions in digital communication and interactions, for example micro politics of emotions [40], or interaction rituals [41].

### 5.2 Digital Followership

Avolio et al point out that not enough attention is being paid to followers and their ability to influence and shape leadership (see Table 1) [18]. For example, mobile devices and ESN can be used as a means of leadership emerging from followers. Sørensen also points out that IS research needs to expands its field of analysis to, for example, mobile devices or ESN, as opposed to the traditional focus on one major technological change in one organization [24]. Mobile devices and ESN allows followers to create new networks organized around knowledge creation or competencies [13].

### 5.3 Generations and Digital Leadership

The empirical data used in the articles analysed in our literature review shows that the age segments addressed vary from between 46 and 65 in 73% of the cases [29]; to 46.2 years of age [31], between 36 and 55 years [32], or 49 years [34]. Future research in digital leadership could analyse if there are any generational differences when collaborating and communicating digitally or face to face, or at least account for the age differences in their reflections. As Avolio et al [18] and Bass [23] discuss, there are mindset differences between generations that could also pressure the leadership system, but IS research so far doesn't seem to address this issue.

### 5.4 Gender in Digital Leadership

In our analysis it was easy to see that in the three instances where the gender was mentioned, the findings were based on a disproportionate number of male respondents [29] [31] [32], whereas in the other papers, the gender was not mentioned. This points out for an opportunity in future lines of research to investigate how male and female leaders choose digital technology, and how they are shaped as leaders as a result of their followers' perception, as well as the role of digital communication in altering this perception.

Avolio et al [4] point out that leadership is highly influenced by how followership perceives and construct their view of the leaders in terms of personality, behaviours, and effectiveness and suggest that male and women leaders are perceived differently when it comes to their perceived performance and effectiveness. On the other hand, they suggest that male and female leaders appropriate technology differently: males based on how useful it is, and females based on ease of use and subjective norms of interactions.

### 5.5 Culture and digital leadership

In the demographics it was observed that there was a high focus on findings based on data collected in the US, however, the relation between culture and the findings was not discussed. Avolio et al [4] point out that different cultures might need different leadership style and provide the example of paternalistic leadership as being predominant for example in China, whereas it could not be applicable in for example the Scandinavian culture. Furthermore, remote work allows for multiple cultures to collaborate. We encourage future research on digital leadership to address culture as a contingency, and to discuss digital leadership practices in multicultural digital teams.

# 6      Conclusion

Leadership is changing as a result of technology, the changing nature of work, organizations, and humans, as well as due to the increased usage of digital communication. As a result, new forms of leadership are needed, which we name in this paper "digital leadership". Our focus is to inquire the IS perspective on digital leadership as well as analyse how the IS research to date can inform future research in digital leadership.

Due to the novelty of digital leadership, we turn to e-leadership. E-leadership is a field emerged from leadership and psychology. This field inspires our definition of digital leadership as well as the use of a theoretical lens to guide our discussion on digital leadership within IS.

We define digital leadership as a process of social influence mediated by technology to produce a change in attitudes, feelings, thinking, behaviour and/or performance with individuals, groups, and/or organizations, which can occur at any hierarchical level in an organization and can involve one-to-one and/or one-to-many interactions. Digital leadership can be analysed at a micro, macro, or meso level.

The micro level is concerned with analysing leaders, followers, leader-follower dyads, context, or the collective, from the perspective of traits (how one is), behaviours (what one does), cognitions (what and how one thinks), and emotions (what one feels). Future research could analyse how these aforementioned perspectives of looking at leadership at a micro level are overlapping with the terminology that IS research has been found to use when discussing leadership: roles, skills, responsibilities, competencies and styles.

The macro level looks into strategic leadership (transactional and transformational leadership), strategic alignment, strategic competencies, leadership contingency, and exploration/exploitation strategies.

The meso level looks into how leadership is changing due to digitalization, business transformation, and the need to create value with IT.

From applying this theoretical lens, we find that IS research has an exclusive focus on the leader's traits, cognitions, and behaviours; leaving out emotions, followers, or the collective, and only briefly addressing dyads or the context. Although IS research focuses on the leader, the leader analysed can be described as a male with the age between 36-55, occupying a position of a director, top manager, or CIO, and located in the US. The theoretical lens helps IS research see the gaps that can be addressed in future research. At the same time, we find that IS research has a strong focus on the macro and the meso level, and we find a broader spectrum of issues than e-leadership proposes at these levels, which we have mentioned few paragraphs before when describing the meso and the macro level.

Beyond a clear invitation for IS research to focus on the gaps found by applying the theoretical lens of e-leadership, as well as to capitalize on its strengths mentioned above, we invite future paths of research in digital leadership to also be connected to emotions, followership, generations, gender, and culture.

## 7       Limitations

Our study has a number of limitations. Avolio et al's framework [18] is built at least partly on face-to-face leadership assumptions, digital leadership requires new theories, future research would profit on approaching digital leadership with empirical methods, as opposed to grounding them in previous leadership theories. Furthermore as digital leadership is a multidisciplinary field, there is a need to look outside IS, in related fields such as psychology, leadership, neuroleadership, or management. Furthermore, in a follow-up article we wish to use a different coding methodology supported by a qualitative analysis software, and inquire practitioners' understanding of digital leadership.

**References**

1. Kotter, J., P.: Leading Change. Business Essentials, München (2011).
2. Orlikowski, W. J.: The Duality of Technology: Rethinking the Concept of Technology in Organizations. Organization Science, 3 (3) pp. 398–427 (1992).
3. DeSanctis, G., Poole, M., S.: Capturing the Complexity in Advanced Technology Use: Adaptive Structuration Theory. Organization Science, 5(2), pp. 121-147 (1994).
4. Avolio, B.J., Kahai, S., Dodge, G., E.: E-Leadership. The Leadership Quarterly, 11 (4), pp. 615–668 (2000).
5. Eberly, M. B., Johnson, M., D., Hernandez, M.: An Integrative Process Model of Leadership: Examining Loci, Mechanisms, and Event Cycles. American Psychologist, 68(6), pp. 427–443 (2013).
6. Hernandez, M., Eberly, M., B., Avolio, B., J.: The Loci and Mechanisms of Leadership: Exploring a More Comprehensive View of Leadership Theory. The Leadership Quarterly, 22(6), pp. 1165–1185 (2011).
7. El Sawy, O., A., Amsinck, H., Kræmmergaard, P., Vinther, A., L.,: Building the foundations and enterprise capabilities for digital leadership: the LEGO experience. Society for Information Management. (2015)
8. Peppard, J., Ward, J.: Beyond strategic information systems: Towards an IS capability, The Journal of Strategic Information Systems 13(2), pp. 167–194 (2004).
9. Davenport, T.H., Kirby, J.: Beyond Automation. Harvard Business Review. (2015).
10. Zammuto, R. F., Griffith, T., L., Majchrzak, A.: Information Technology and the Changing Fabric of Organization. Organization Science, 18 (5), pp. 749–762 (2007).
13

issues than e-leadership proposes at these levels, which we have mentioned few paragraphs before when describing the meso and the macro level.

Beyond a clear invitation for IS research to focus on the gaps found by applying the theoretical lens of e-leadership, as well as to capitalize on its strengths mentioned above, we invite future paths of research in digital leadership to also be connected to emotions, followership, generations, gender, and culture.

## 7       Limitations

Our study has a number of limitations. Avolio et al's framework [18] is built at least partly on face-to-face leadership assumptions, digital leadership requires new theories, future research would profit on approaching digital leadership with empirical methods, as opposed to grounding them in previous leadership theories. Furthermore as digital leadership is a multidisciplinary field, there is a need to look outside IS, in related fields such as psychology, leadership, neuroleadership, or management. Furthermore, in a follow-up article we wish to use a different coding methodology supported by a qualitative analysis software, and inquire practitioners' understanding of digital leadership.

**References**

1. Kotter, J., P.: Leading Change. Business Essentials, München (2011).
2. Orlikowski, W. J.: The Duality of Technology: Rethinking the Concept of Technology in Organizations. Organization Science, 3 (3) pp. 398–427 (1992).
3. DeSanctis, G., Poole, M., S.: Capturing the Complexity in Advanced Technology Use: Adaptive Structuration Theory. Organization Science, 5(2), pp. 121-147 (1994).
4. Avolio, B.J., Kahai, S., Dodge, G., E.: E-Leadership. The Leadership Quarterly, 11 (4), pp. 615–668 (2000).
5. Eberly, M. B., Johnson, M., D., Hernandez, M.: An Integrative Process Model of Leadership: Examining Loci, Mechanisms, and Event Cycles. American Psychologist, 68(6), pp. 427–443 (2013).
6. Hernandez, M., Eberly, M., B., Avolio, B., J.: The Loci and Mechanisms of Leadership: Exploring a More Comprehensive View of Leadership Theory. The Leadership Quarterly, 22(6), pp. 1165–1185 (2011).
7. El Sawy, O., A., Amsinck, H., Kræmmergaard, P., Vinther, A., L.,: Building the foundations and enterprise capabilities for digital leadership: the LEGO experience. Society for Information Management. (2015)
8. Peppard, J., Ward, J.: Beyond strategic information systems: Towards an IS capability, The Journal of Strategic Information Systems 13(2), pp. 167–194 (2004).
9. Davenport, T.H., Kirby, J.: Beyond Automation. Harvard Business Review. (2015).
10. Zammuto, R. F., Griffith, T., L., Majchrzak, A.: Information Technology and the Changing Fabric of Organization. Organization Science, 18 (5), pp. 749–762 (2007).